# Secure Open Federation of IoT Platforms Through Interledger Technologies - The SOFIE Approach

Dmitrij Lagutin[1], Francesco Bellesini[3], Tommaso Bragatto[4], Alessio Cavadenti[4], Vincenzo Croce[5], Yki Kortesniemi[2], Helen C. Leligou[6], Yannis Oikonomidis[6], George C. Polyzos[7], Giuseppe Raveduto[5], Francesca Santori[4], Panagiotis Trakadas[6], Matteo Verber[5]

[1]Department of Communications and Networking, [2]Department of Computer Science, Aalto University, Espoo, Finland.
[3]Emotion s.r.l., Bastia, Italy. [4]ASM Terni, Terni, Italy. [5] Engineering Ingegneria Informatica S.p.A., Rome, Italy.
[6]Synelixis Solutions, Athens, Greece.
[7]Mobile Multimedia Laboratory, Athens University of Economics and Business, Athens, Greece.

*Abstract* — *The lack of interoperability among IoT platforms has led to a fragmented environment, where the users and society as a whole suffer from lock-ins, lack of privacy, and reduced functionality. This paper presents SOFIE, a solution for federating the existing IoT platforms in an open and secure manner using Distributed Ledger Technologies (DLTs) and without requiring modifications to the IoT platforms, and describes how SOFIE is used to enable two complex real life pilots: food supply chain tracking from field to fork and electricity distribution grid balancing with guided electrical vehicle (EV) charging. SOFIE's main contribution is to provide interoperability between IoT systems while also enabling new functionality and business models.*

*Keywords* — *Internet of Things; distributed ledger technologies (DLTs); blockchains; smart contracts; smart grid; supply-chain tracking; food provenance*

## I. Introduction

Fragmentation and lack of interoperability among different platforms is a major issue with Internet of Things (IoT). Currently, IoT platforms and systems are vertically oriented silos unable (or unwilling) to exchange data with, or perform actions across, each other. This leads to multiple problems: reduced competition and vendor lock-ins, as it is difficult for customers to switch IoT providers, worse privacy as vendors usually force their customers to move at least some of their data or metadata to the vendor's cloud, and reduced functionality compared to what better interoperability would afford. Since IoT systems are becoming prevalent in everyday life, lack of interoperability and limited use of relevant data is growing into a significant problem for the whole society.

The EU H2020 project SOFIE [1] enables applications to utilise heterogeneous IoT platforms and autonomous things across technological, organisational, and administrative borders in an open and secure manner, thus simplifying the reuse of existing infrastructure and data, and allowing the creation of open business platforms. This goal is accomplished through distributed ledger technologies (DLTs) [2] and without requiring modifications to the existing IoT platforms. In the long term this will also enable open data markets, where participants can buy and sell IoT data and access to IoT actuation (or more generally: dictate rules for access to data and actuation) in a decentralized and automated manner.

The contributions of this paper include: 1) a description of how Distributed Ledger Technologies (DLTs) and interledger techniques can be utilized to enable secure and open federation among heterogeneous IoT platforms and 2) two examples of how such federation can be leveraged in complex real-world systems, namely food supply chain and electricity distribution grid balancing with electrical vehicle (EV) charging.

The paper is organized as follows. Section 2 describes DLTs and their applicability to the IoT interoperability problem, and Section 3 describes related work, both in IoT and the DLT area. Then, Section 4 describes the SOFIE federation approach, and Sections 5 and 6 detail how the approach is leveraged in real-world use cases. Section 7 provides a discussion of forward-looking applications and more elaborate extensions of the use cases leading to new business models and secure open data markets. Finally, Section 8 concludes the paper and describes our ongoing and future work.

## II. Distributed Ledger Technologies

Distributed Ledger Technologies (DLTs) such as blockchains offer decentralized solutions for collaboration and interoperability. One of the main features of DLTs is the immutability of data: ledgers are append-only databases where existing data cannot be modified and only new data can be added. Another major feature of DLTs is a distributed consensus mechanism [3], which controls what and how data is added to the ledger. Finally, DLTs also replicate data to participating nodes thus improving availability. Because of these three properties, DLTs avoid a single point of failure and offer resilience against many attacks. It is easy to determine if any of the participating nodes in the DLT are misbehaving and even in an extreme case where an attacker manages to control the majority of the DLT's resources, the attacker still cannot modify the existing data, only control the addition of new data. Technically, DLTs can be implemented with different levels of openness. They can be fully *open* (permissionless), which means that anyone can join the DLT and propose transactions; most well-known DLTs such as Bitcoin and Ethereum are based on this principle. However, DLTs can also be permissioned such as *semi-open*, in which case read access is open to everyone but write access is restricted, or *closed*, in which case both read and write access are restricted.

The main practical innovation of DLTs is the enablement of distributed trust. While there have been multiple proposals for distributed databases in the past, they have mostly concentrated on the distributed implementation, while the trust model has remained firmly centralized. In contrast, DLTs allow various entities, such as individuals, organizations, and companies, which may not fully trust each other, to collaborate in a safe and transparent manner, with only a low risk of being cheated by others. This makes DLTs a natural approach for solving the interoperability problem between IoT platforms.

There exists a large number of DLTs each offering different trade-offs in terms of latency, throughput, consensus algorithm, etc. thus rendering them suitable for different types of applications. In complex systems it is therefore often not feasible to use only a single DLT for everything, hence the interledger approach that allows different DLTs to exchange data with each other is required in many situations. Using multiple ledgers is also beneficial for privacy reasons. Participants within a DLT need to be able to access all data stored in that DLT to independently verify its integrity, which encourages the participants to use private ledgers, and store only a subset of the data to the main ledger used for collaboration with others. Multiple ledgers are also necessary crypto-agility as cryptographic algorithms used by DLTs (such as SHA-256) will not stay safe forever, thus it is necessary to have a mechanism to transfer data from one ledger to another.

Previously, multiple interledger approaches have been used, and while there are no established standards for interconnecting DLTs, a few repeating patterns have been observed: a shared motivation for the interledger solutions is to move away from the 'one chain to rule them all' model to one that allows the interconnection of multiple ledgers. Interledger approaches include: 1) atomic cross-chain transactions, 2) sidechains, 3) bridging, 4) transactions across a network of payment channels, 5) ledger-of-ledgers, and 6) the W3C Interledger Protocol (ILP). Voulgaris et al. compare the approaches according to whether they support the transfer or the exchange of value, their interconnection trust mechanism, complexity, scalability, and cost [4].

A concrete example of the use of interledger techniques in an application could be the following: Some entities (from different companies, perhaps operating in the same market) decide to use the permissioned Hyperledger Fabric blockchain, which provides no-cost transactions and chaincode for transaction automation. Entities also decide to use Ethereum in order to make payments and fully automate the whole process with smart contracts for transactions involving payments. An interledger mechanism can be used to interconnect the two ledgers in a way that ensures atomic transactions, i.e., payments are made only for complete, successful transactions and vice-versa, transactions do not complete successfully unless payment is made, and cannot be retracted unilaterally.

### III. RELATED WORK

There exist several proposals for solving the IoT interoperability problem. Some approaches rely on creating a new interoperability layer, which is not feasible in most cases, since it requires making changes to the existing IoT platforms.

Other approaches, including BIG IoT [5], aim to allow interoperability between IoT systems through an API and Marketplace; however the proposed marketplace is designed to be centralized, limiting its use. WAVE [6] provides a decentralized authorization solution for IoT devices using a private Ethereum blockchain and smart contracts, but it assumes that all IoT devices are able to interact with the blockchain, which is not feasible for many constrained devices.

There are also application specific approaches utilizing DLTs for, e.g. energy trading [7][8][9]. Often they use cryptotokens issued by a single party as currency, which can lead to speculation. Such an approach distances the solution from its actual use case, and while the cryptocurrency was the original use case of blockchains, it is important to use separate DLTs for performing payments and for other uses, such as asset tracking, logging, etc.

Therefore, the existing work does not fully address the need for an open, decentralized solution for the IoT interoperability problem, which supports existing IoT platforms and enables new open business models.

### IV. THE SOFIE FEDERATION APPROACH

The main aim of SOFIE is to federate existing IoT platforms in an open and secure manner, without making internal changes to the platforms themselves. Here, openness refers both to technical aspects (interfaces, implementation, etc.) and to flexible and administratively open business models. The approach also aims to preserve users' privacy and be compliant with the EU General Data Protection Regulation (GDPR), which requires the minimisation of personal data collection. One promising technology utilized by SOFIE for improving privacy are decentralized identifiers (DIDs) [10][11] that allow users to create and control short-lived digital identifiers in a flexible manner.

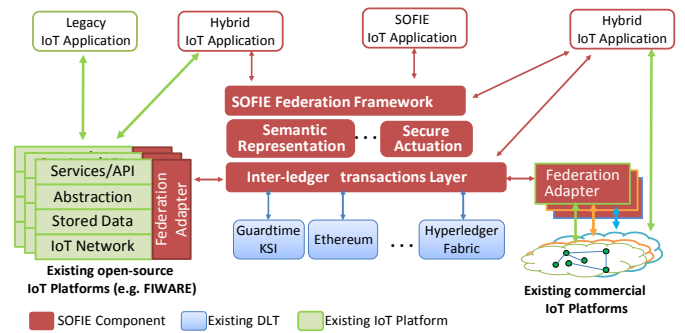

Fig. 1. An overview of the SOFIE architecture

An overview of the SOFIE architecture is depicted in Fig. 1. At the core of SOFIE is the interledger transactions layer, which allows transactions between multiple ledgers such as Ethereum, Hyperledger Fabric, Guardtime KSI, etc. [4]. Federation Adapters are used to allow interactions between the existing IoT platforms and the SOFIE federation framework without any changes to the IoT platforms themselves. SOFIE will also utilize existing work in this area, such as W3C Web of Things (W3C) and the FIWARE IoT platform. As an example, DLTs and smart contracts can be used for managing access control and performing automatic payments even with constrained and disconnected IoT devices [12][13]. The first version of the SOFIE federation framework will be released under open source license in autumn 2019.

The SOFIE approach will be demonstrated in four real-life pilots within three different topic areas: 1) agricultural supply chain, where produce growth and transportation conditions are tracked from field to fork, 2) power balancing of the electrical grid by offering incentives to EV owners to charge their cars at certain times and perhaps locations, 3) mixed reality mobile gaming where gamers can interact with the real world through IoT devices, and 4) utilizing data from electricity smart meters to develop various applications, e.g., to suggest the best electricity provider for a given user profile. The first two pilots will be described in more detail in subsequent sections.

## V. TRACKING FOOD FROM FARM TO FORK

The farm-to-fork pilot demonstrates the application of SOFIE to a community-supported and heterogeneous end-to-end agricultural food chain scenario. Contrary to traditional centralised food supply chain systems, it leverages blockchain technology to ensure transparency in data management, thus increasing the level of traceability and integrity of data without the need for a centralized authority.

By using smart contracts, transactions over heterogeneous IoT ecosystems (segments of food supply chain as described below) are automated, thus reducing the chances of fraud, cutting out corresponding mediation expenses and transaction costs, and providing immutable proofs of interactions between different parties. In view of this, consumers can now reliably verify the provenance of a specific product from farm to fork. This gives consumers the ability to make decisions about their food based on health and ethical concerns, including environmental sustainability, fair labour practices, the use of fertilizers and pesticides, and other similar issues. On the producers' side, they will be able to launch new products with a description, pricing, quantity and photos, while customers may interact with the marketplace, looking for products that fulfil certain requirements or preferences. This would not be feasible if the various DLTs were not able to exchange information and if this information was not semantically annotated so that it can be appropriately searched.

The path from farm to fork is split into 5 segments as depicted in Fig. 2.

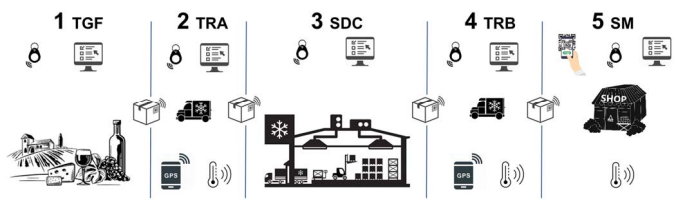

Fig. 2. An overview of the SOFIE food-chain pilot, describing how produce moves from the farm to the supermarket through transporters and distributers

**Smart Farm (SF)**: In the farm, there are multiple sensor nodes capable of measuring parameters such as temperature, humidity, wind speed/direction, rainfall, and soil moisture. The data from all these sensors are stored in the Ethereum-based Smart Farm IoT platform.

**Transportation Route A (TRA)**: This segment covers the path from SF to the Storage & Distribution Centre (SDC). The vehicles are equipped with GPS and temperature sensors, and all data from those sensors will be stored in the Ethereum-based Transportation IoT platform.

**Storage & Distribution Centre (SDC)**: SDC is where the smart boxes with farm crops will be stored until they are transported to the Supermarket. In SDC, a number of sensors monitor, among other parameters, temperature and presence of the boxes. The data from these sensors are stored in the Hyperledger Fabric based SDC IoT Platform.

**Transportation Route B (TRB)**: This segment covers the path from SDC to Supermarket and the vehicles are again equipped with GPS as well as temperature sensors. Data from those sensors is stored in the Ethereum-based Transportation IoT platform.

**Supermarket (SM)**: SM is the products' final destination in the pilot's context. There are two areas of interest in the SM: the storage area, where the boxes are kept until they are placed in the customer area, and the customer area, where the products are available for the customers. The collected information about the environmental conditions where products are stored and later exhibited, will be stored in a Hyperledger Fabric based platform. Before the products are removed from the smart boxes to be placed to the customers' area, QR labels are created and applied on the crop packages so that SM customers can retrieve this information using their smartphones.

To enable the end-to-end chain, the different distributed ledgers used by each of the segments are liaised through a Consortium Ledger (Ethereum) instance that allows interledger communication. The Consortium Ledger is run in a distributed manner by members of the consortium and supervised by a Legal Entity on a national or European level (association or public authority). In practice, the role of the supervisor is to control only the membership of blockchain nodes. For example, in Hyperledger Fabric terminology, to enact just the "certification authority" role.

Here, the main advantage of SOFIE's architectural approach compared to other approaches is that it is agnostic to the technology and technical specifications of the integrated IoT environments, e.g. segments of the food supply chain, and that it provides an easy to use and non-disruptive solution to federation by introducing a transparent data adaptation layer that enables interoperability over the datasets generated and processed in different distributed ledgers.

## VI. GRID BALANCING WITH ELECTRICAL VEHICLES

In a second pilot, the SOFIE platform is utilised to balance a real energy network, namely the distribution grid of the city of Terni located in central Italy. There, a notable amount of energy is produced locally by distributed photovoltaic plants [14], which on occasion can cause Reverse Power Flow, when unbalances between produced and consumed energy occur. To avoid this abnormal operation [15][16], electrical vehicles (EVs) will be offered significant incentives to match their EV charging needs with the distribution network's requirements.

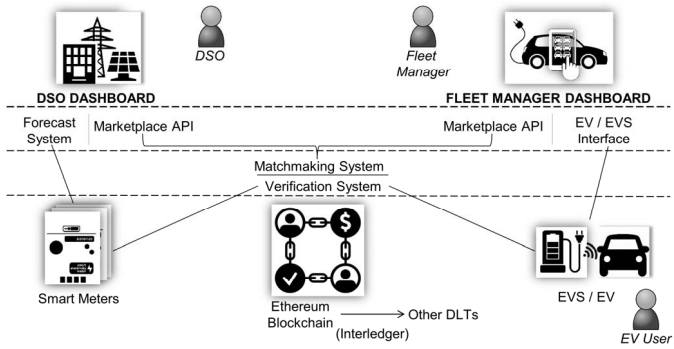

Fig. 3. An overview of the SOFIE energy pilot, describing how DSO, EV fleet manager, and EV users use decentralized marketplace to optimize the load on electrical grid

The actors in the pilot, as depicted in Fig. 3, are the *Distribution System Operator (DSO)*, who is responsible for grid management, the *Fleet Manager*, who is the manager of the charging stations and represents EVs in energy price negotiations, and the *EV users*, who receive information and requests about the optimal scheduling of the charging of their vehicle. The SOFIE platform is then utilised to run a decentralised marketplace enabling a peer to peer exchange mechanism between DSO and fleet manager, thus forming an end-to-end scenario from production via distribution to storage and consumption.

The DSO and the fleet manager interact with the system through their dedicated dashboards that show near real-time data collected from the two IoT subsystems (i.e. smart meters for the DSO and EV/EVS sensors for the fleet manager). The actors create market requests and offers accordingly. The business logic for the requests and offers collection and for the winning offer selection algorithm is coded in smart contracts, ensuring transparency and auditability of the whole process.

The current version of the smart contract implements an auction mechanism, in which the best offer is selected following the "lowest bidder" rule. In the future, the upgraded version of the smart contract will consider a different matchmaking algorithm, based on the clearing price algorithm used in commodity trades. In addition, the smart meter readings are stored on blockchain to ensure transparency, and the blockchain will also contain data of electric vehicles, charging stations, and charging events. Such data will be used for payments by the DSO to the fleet manager and for rewarding the users (through tokens or discounts) in an automated manner.

Two main scenarios will be studied in the pilot, called *day ahead planning* and *intraday contingency planning* scenario, respectively. In the *day ahead planning* scenario the DSO needs to shave peaks of locally produced energy, so it will put a flexibility request in the day-ahead market asking for an amount of energy to be drawn at specific time intervals and location and provides incentives (e.g. tokens or discounts). Meanwhile, the fleet manager matches fleet needs with offers in the platform to achieve the maximum bonus available in terms of incentives, placing the corresponding offers in the marketplace. On the other hand, in the *intraday contingency planning* scenario the DSO puts out a flexibility request asking for an amount of energy (kWh), a timeslot, and a location (GPS coordinates) while providing an incentive (tokens) in order to shave peaks of locally produced power in the same day. The marketplace will then automatically identify potential candidates to fulfil the request based on user type, current location, residual autonomy, and EV's current status. The selected EV users will receive a direct notification, offering a token incentive if they agree to charge the vehicle using an assigned charging station in a specific time interval.

The scenarios can be easily extended to include a *retailer* actor in charge of accounting, providing benefits to the two main actors involved: the DSO benefits of the grid stability provided and the fleet manager can reduce the overall charging costs to be paid to the retailer thanks to the incentives awarded by the DSO.

The marketplace will operate on a private Ethereum based Blockchain, granting privacy (i.e., data cannot be read by external parties) and reducing transactions costs and times (i.e., mining is not required). Thanks to the interledger layer provided by SOFIE, this "first layer" will be paired with a public DLT acting as a "second layer", where the status of the private blockchain is periodically synchronized, granting security and auditability, thus protecting the data stored in the first layer DLT from any alterations. The interledger capabilities theoretically will also permit seamless access to the decentralized energy marketplace to fleet managers or DSOs already using their own existing blockchain solution.

Compared to existing solutions, a key benefit of SOFIE is the federation of existing platforms (i.e. the EV platform and the Advanced Metering Infrastructure managed by the DSO). Thanks to the technology-agnostic SOFIE federation framework, DSO and fleet manager can interoperate in the same decentralized marketplace, keeping their own internal IoT platforms unmodified. SOFIE implementation is producing benefits for both sides, on the one hand, services are temporally synchronized and benefits of the actors are optimized, on the other hand, new services can be investigated in the future and provided to third parties, opening new business models for the actors themselves e.g. a stable prediction of the active power exchanges with the Transmission System Operator (TSO), carried out by balancing the loads of the charging points, can be subject to monetary incentives.

## VII. DISCUSSION

Key ingredients of the SOFIE approach to IoT interoperability are the exploitation of multiple, appropriate distributed ledgers and interledger technologies to support openness, decentralization, trust, security, privacy, automation, and auditability.

Openness is important mostly from an economic and business perspective, as it allows any player to freely participate in the ecosystem just by following the set rules and without one or a few players having veto power or controlling the ecosystem. Some DLTs even rely on distributed governance thus allowing the evolution of the rules from within the system. Openness may be undesirable for (myopic) individual players, but it is beneficial and perhaps even critical for society as a whole. Decentralization is a key ingredient for openness, but it also supports robustness, avoiding single or few points of failure, which would also be targets for attacks.

Building and propagating trust is important not only for efficient economic activities but also for usability. Security is of critical importance: not only is it required at the individual IoT system level, but our focus here is on end-to-end and at the level of the whole system, including the interfacing mechanisms and components, which may be even more vulnerable. It includes security against both external and internal attacks with a special important case being attacks from interconnected systems, which need to be provided controlled access. For the IoT, the aspect of security is critical since the IoT is bridging the cyber with the physical world, therefore security breaches can lead to major safety issues.

It is essential that such interconnected systems provide various privacy properties and guarantees, which need to be different at various parts of the world and for different actors. Public DLTs significantly complicate privacy since not only do all parties have access to all information, but replication makes it easy for all to access the information, and immutability leads to availability of all past information that facilitates data correlation and mining, thus revealing even information not directly disclosed. On the other hand, what is bad for privacy is sometimes good for verifiability and auditability.

Perhaps the most important feature of all is automation in a reliable, available, secure, and decentralized manner – and SOFIE utilises smart contracts to fulfil this role. For instance, in order to support openness and privacy in access to data and actuation, an automatic process is required to control the access, perhaps complemented by an associated payment mechanism (which can be provided by cryptocurrencies).

As the pilots demonstrate, SOFIE allows the integration of different IoT platforms with Federation Adapters thus requiring no changes to the platforms. The resultant systems can then benefit from the increased functionality and privacy provided by the federation approach. The four real-life pilots of SOFIE also enable interesting cross-pilot interactions. For example, a mobile gamer can receive in-game assets as reward for buying ethically produced agricultural produce, or the possession of a certain in-game asset could offer discounts for electrical vehicle charging. All pilots are currently being implemented and their results will be presented in future publications.

## VIII. CONCLUSIONS

This paper describes how SOFIE utilises Distributed Ledger Technologies (DLTs) for providing interoperability between IoT platforms in an open and secure manner. This work has shown that using DLTs allows more flexible co-operation between different parties in multiple use cases, such as food supply chain and electricity grid load balancing. The SOFIE solution is tested in four real-life pilots, which will also enable interesting cross-pilot interactions.

In the longer term, this approach will also enable open data markets and allow the creation of new business models around IoT data.

ACKNOWLEDGMENT

This research has received funding from European Union's Horizon 2020 research and innovation programme under grant agreement No. 779984, and from Business Finland under grant No. 3387/31/2017.